\documentclass[submission,copyright,creativecommons]{eptcs}
\providecommand{\event}{DSL 2011}

\usepackage{listings}
\lstMakeShortInline|
\lstnewenvironment{code}{}{}

\newtheorem{Exercise}{Exercise} 
\newcommand\aside[1]{}

\title{Implementing Explicit and Finding Implicit Sharing in Embedded DSLs}
\author{Oleg Kiselyov
\email{oleg@okmij.org}}
\def\titlerunning{Sharing in embedded DSLs}
\def\authorrunning{Oleg Kiselyov}

\begin{document}
\maketitle

\begin{abstract}   
Aliasing, or sharing, is prominent in many domains, denoting that two
differently-named objects are in fact identical: a change in one
object (memory cell, circuit terminal, disk block) is instantly   
reflected in the other.  Languages for modelling such domains should
let the programmer explicitly define the sharing among objects or
expressions. A DSL compiler may find other identical expressions and
share them, implicitly. Such common subexpression elimination is
crucial to the efficient implementation of DSLs. Sharing is tricky in
embedded DSL, since host aliasing may correspond to copying of the
underlying objects rather than their sharing.
  
This tutorial summarizes discussions of implementing sharing in
Haskell DSLs for automotive embedded systems and hardware description 
languages. The technique has since been used in a Haskell SAT solver
and the DSL for music synthesis. We demonstrate the embedding in pure
Haskell of a simple DSL with a language form for explicit
sharing. The DSL also has implicit sharing, implemented via
hash-consing. Explicit sharing greatly speeds up hash-consing. The
seemingly imperative nature of hash-consing is hidden beneath a simple
combinator language. The overall implementation remains pure
functional and easy to reason about.
\end{abstract}

\begin{quote}
I think all DSLs suffer from the same problems:
sharing and recursion. I've used wrappers for CSound, SuperCollider,
MetaPost, they all have these problems.
\rm Henning Thielemann~\cite{Thielemann}
\end{quote}

\section{Introduction}

We present implicit and explicit sharing in the original context of
embedded domains-specific language (DSL) compilers. The sharing
implementation techniques have since found other uses, e.g., writing SAT
solvers~\cite{funsat}. Embedded compilers~-- typical for circuit description,
embedded control systems or GPU programming DSLs~-- complement the
familiar embeddings of a DSL in a host language as a library or an
interpreter. For example, we may build a circuit model in
Haskell using gate descriptions and combinators provided by the DSL
library. We may test the circuit in Haskell by running the model
on sample inputs. Eventually we have to produce a Verilog
code or a Netlist to manufacture the circuit. Likewise, a control
system DSL program should eventually be compiled into machine code and
burned as firmware. An embedded compiler thus is a host language
program that turns a DSL program into a (lower-level) code.

One of the important tasks of a compiler is the so-called ``common
subexpression elimination''~-- detecting subexpressions denoting the
same computation and arranging for that computation to be performed
once and the results shared. This optimization (significantly)
improves both the running time and compactness of the code and is
particularly important for hardware description and firmware
compilers. As DSL implementers, it becomes our duty to detect
duplicate subexpressions and have them shared. We call this detection
implicit sharing, to contrast with the sharing explicitly declared by
users. Imperative languages, where it matters whether an expression
or its result are duplicated, have to provide a form (some sort of a
local variable introduction) for the programmer to declare the sharing
of expression's result. In the present paper we limit ourselves to
side-effect--free expressions such as arithmetic expressions or
combinational circuits. Explicit sharing is still important: sharing
declarations can significantly reduce the search space for common
subexpressions and prevent exponential explosions. We shall cite
several examples later. Sharing declarations also help human
readers of the program, drawing their attention to the
`common' computations.

A common pitfall in implementing explicit sharing is attempting to use
the |let| form of the host language to express sharing
in the DSL code. \S\ref{s:obj-let} will show that although the |let|-form
may speed up some DSL interpretations, it leads to code
duplication rather than sharing in the compilation result.

Before we get to that discussion, we introduce our running example in
\S\ref{s:cse-problem}, describing an unsuccessful attempt to detect
sharing in the course of implementing real DSLs. Show-stopping was the
expression comparison, which, in pure language is structural and
requires the full traversal of expressions. In \S\ref{s:ptr-cmp} we
review the main approaches to speed-up the comparison, all relying on
some sort of `pointer' equality. Our method of expression comparison
and sharing detection is presented in \S\ref{s:hash-cons}.  The method
is a pure veneer over hash-consing, alleviating the cost of
comparison. Its main benefit is the facilitation of explicit sharing,
described in \S\ref{s:obj-let}.

The code accompanying the paper is available online at
\url{http://okmij.org/ftp/tagless-final/sharing/}.

\section{Detecting sharing: necessity and difficulty}
\label{s:cse-problem}

Our running example is a show-stopping problem encountered by the
implementer of an embedded DSL compiler for an embedded control
system, posed in \cite{Hawkins-CSE}.  The example illustrates the need
and the difficulty of common subexpression elimination. 
The problem was posed for arithmetic expressions, which are part of
nearly every DSL. Typically arithmetic expressions are embedded in
Haskell as the values of the datatype\footnote{%
See the file \url{ExpI.hs} in the accompaniment for the complete code.}
\begin{code}
data Exp
   = Add Exp Exp
   | Variable String
   | Constant Int
   deriving (Eq, Ord, Show)
\end{code}
Here are the sample expressions in our DSL:
\begin{code}
exp_a = Add (Constant 10) (Variable "i1")
exp_b = Add exp_a (Variable "i2")
\end{code}

The implementer wanted to compile these DSL expressions into C or
the machine code, using the standard approach of finding common
subexpressions and sharing them. To make the sharing explicit, the
expression tree is converted into a directed acyclic graph (DAG).  The
graph is then topologically sorted and each subexpression is assigned
a C operation or a machine instruction.

The first step, detecting identical subexpressions, was the most
troublesome. The step is crucial since common subexpressions are
abound, being easy to create. We show two real-life examples, to be used
throughout the paper. The first example is the multiplication by a
known integer constant. Our DSL does not have the multiplication
operation (8-bit CPUs rarely have the needed instruction).
Nevertheless, we can multiply a DSL expression by the known constant
using repeated addition. Here is the standard efficient procedure
based on the recursive subdivision:
\begin{code}
mul :: Int -> Exp -> Exp
mul 0 _ = Constant 0
mul 1 x = x
mul n x | n `mod` 2 == 0 = mul (n `div` 2) (Add x x)
mul n x = Add x (mul (n-1) x)
\end{code}
The result of the sample expression
\begin{code}
exp_mul4 = mul 4 (Variable "i1")
\end{code}
shows two identical subexpressions of adding the variable |i1| to itself:
\begin{code}
Add (Add (Variable "i1") (Variable "i1")) (Add (Variable "i1") (Variable "i1"))
\end{code}
The result of |mul 8 (Variable "i1")| shows twice as much duplication.

The other running example, from the domain of hardware description,
is |sklansky| by Naylor \cite{Naylor-sharing}, with further
credit to Sheeran and Axelsson.  The example computes the running sum
of given expressions; like the previous multiplication example,
|sklansky| uses recursive subdivision to expose more parallelism and
reduce latency:
\begin{code}
sklansky :: (a -> a -> a) -> [a] -> [a]
sklansky f [] = []
sklansky f [x] = [x]
sklansky f xs = left' ++ [f (last left') r | r <- right']
  where
    (left, right) = splitAt (length xs `div` 2) xs
    left'  = sklansky f left
    right' = sklansky f right
\end{code}
The pretty-printed result of |sklansky Add (map (Variable . show) [1..4]|
\begin{code}
["v1","(v1+v2)","((v1+v2)+v3)","((v1+v2)+(v3+v4))"]
\end{code}
demonstrates the triplication of the subexpression |v1+v2|. The
duplication should be eliminated when we build the circuit.

In the process of converting expressions like |exp_mul4| to a DAG,
the implementer \cite{Hawkins-CSE} had to compare subexpressions. It
is there he encountered a problem: in a pure language, we may only
compare datatype values structurally. General pointer comparison
destroys referential transparency and parametricity. To
check that the summands of the top-level addition in |exp_mul4| are
identical, we have to therefore traverse them completely. Such comparisons of
two expression trees take more and more time with larger programs
(say, as we multiply by bigger integers).  ``As
these trees grow in size, the equality comparison in graph
construction quickly becomes the bottleneck for DSL compilation.
What's worse, the phase transition from tractable to intractable is
very sharp.  In one of my DSL programs, I made a seemingly small
change, and compilation time went from milliseconds to
not-in-a-million-years.''\cite{Hawkins-CSE}.  His message, entitled
``I love purity, but it's killing me'' was a cry for help: he was
about to give up on Haskell.  He wondered how to dramatically speed-up
comparisons, or find a better method of detecting common
subexpressions and sharing them.

\section{Pointer comparison}
\label{s:ptr-cmp}

Sharing detection, specifically, fast identity comparison of
expressions, is a common metaprogramming problem and has been
investigated extensively. This section reviews the main approaches,
which are all based on some sort of `pointer' equality. They associate
with an expression a unique datum, e.g., an integer, admitting
efficient comparison. For instance, we may define the data type of
labeled expressions, where each variant carries a unique integer
label\footnote{%
The complete code for this section is in the file
  \url{Ptrs.hs}.}:
\begin{code}
type Lab = Int
data ExpL
   = AddL Lab ExpL ExpL
   | VariableL Lab String
   | ConstantL Lab Int
   deriving Show
\end{code}
|ExpL| expressions are fast to compare, by comparing their
labels. The full traversal of expressions is no longer needed:
\begin{code}
instance Eq ExpL where
    e1 == e2 = label e1 == label e2
     where
     label (AddL p _ _)    = p
     label (ConstantL p _) = p
     label (VariableL p _) = p
\end{code}

The first approach to building labeled expressions is manual
labeling. For example, we construct our sample expressions as follows,
taking great care to pick unique labels:
\begin{code}
expL_a = AddL 3 (ConstantL 1 10) (VariableL 2 "i1")
expL_b = AddL 4 expL_a (VariableL 5 "i2")
\end{code}
Needless to say this approach is greatly error-prone, let alone
tedious. When implementing the |mul| example we stumble on another
complication: the manual threading of a counter to generate unique labels.
Some sort of automation is direly needed.

A promising and increasingly popular way to hide and automate label
assignment is Template Haskell (see the thesis \cite{acosta-hardware}
for the extensive discussion). A DSL implemented in Template
Haskell quotations can hardly be called `embedded' however. Another
approach is the State monad hiding the counter used for the generation
of unique labels:
\begin{code}
type ExpM = State Lab ExpL

new_labelM :: State Lab Lab
new_labelM = do
  p <- get
  put (p+1)
  return p

run_expM :: ExpM -> ExpL
run_expM m = evalState m 0
\end{code}
The computation |new_labelM|, or `gensym', yields a
unique label; the function |run_expM| runs the monadic computation returning
the produced labeled expression.

To hide the labeling of expression's constructors, we build
expressions with `constructor functions', often called `smart
constructors':
\begin{code}
varM :: String -> ExpM
varM v   = new_labelM >>= \p -> return $ VariableL p v

constM :: Int -> ExpM
constM x = new_labelM >>= \p -> return $ ConstantL p x

addM :: ExpL -> ExpL -> ExpM
addM x y = new_labelM >>= \p -> return $ AddL p x y
\end{code}

The sample expression looks awful
\begin{code}
expM_a = do 
  xv <- constM 10 
  yv <- varM "i1"
  addM xv yv
\end{code}
although appropriate combinators may improve the appearance. In fact,
the multiplication example below is quite perspicuous: the labeling is
well-hidden (compare with |mul| in \S\ref{s:cse-problem}). The
occasional |return| and the `call-by-value application'
|(=<<)| betray the effectful computation:
\begin{code}
mulM :: Int -> ExpL -> ExpM
mulM 0 _ = constM 0
mulM 1 x = return x
mulM n x | n `mod` 2 == 0 = mulM (n `div` 2) =<< addM x x
mulM n x = addM x =<< mulM (n-1) x
\end{code}
The subexpression |addM x x| on the last-but-one line builds 
the addition expression from two identically named summands. The
two occurrences of the Haskell variable |x| denote the same |ExpL|
expression, with the same label. These identical summands are thus
easily identifiable as shared. Indeed, the multiplication-by-4
computation
\begin{code}
expM_mul4 = mulM 4 =<< varM "i1"
\end{code}
when run, yields the labeled expression
\begin{code}
AddL 2 
  (AddL 1 (VariableL 0 "i1") (VariableL 0 "i1")) 
  (AddL 1 (VariableL 0 "i1") (VariableL 0 "i1"))
\end{code}
with the clearly seen sharing: just look at the labels.

\begin{Exercise}\label{ex:addM}
Why we did not define |addM| as follows?
\begin{code}
addM :: ExpM -> ExpM -> ExpM
addM x y = do
  xv <- x
  yv <- y
  p  <- new_labelM
  return $ AddL p xv yv
\end{code}
After all, it does let us write the sample expression |expM_a| 
concisely:
\begin{code}
expM_a = addM (constM 10) (varM "i1")
\end{code}
Hint: implement |mulM| and see the result of |expM_mul4|.
\end{Exercise}

The monadic approach may seem adequate~-- to us, DSL implementers. It
does not however feel right to our users, domain experts. They are
accustomed to writing and manipulating arithmetic expressions as
familiar mathematical objects. Now they have to deal with effectful
computations. O'Donnell, summarizing the dissatisfaction of hardware
designers with such a monadic DSL, wrote: ``A more severe problem is
that the circuit specification is no longer a system of simultaneous
equations, which can be manipulated formally just by 'substituting
equals for equals'. Instead, the specification is now a sequence of
computations that~-- when executed~-- will yield the desired
circuit. It feels like writing an imperative program to draw a
circuit, instead of defining the circuit
directly."\cite{ODonnell-dagstuhl}. The hardware description DSL Lava
has tried the monadic approach and abandoned it.

The most popular approach to detect sharing is a so-called `observable
sharing' \cite{claessen-observable}, which hides the labeling further,
to the point of breaking Haskell and using internal, unsafe operations
of GHC. Fresh label generation is such a benign effect. The
breaking of the referential transparency is hardly noticeable, one
may argue \cite{claessen-observable,Naylor-obs-sharing}. There are
many variations of observable sharing
\cite{acosta-hardware,claessen-observable,Naylor-obs-sharing}: some
rely on |IORef| cells as labels (they can be compared efficiently as
pointers), some use gensym. We demonstrate the latter, and define
gensym, or |new_label|, as a supposedly pure, ordinary Haskell
function:
\begin{code}
{-# NOINLINE counter #-}
counter = unsafePerformIO (newIORef 0)
new_label :: () -> Int
new_label () = unsafePerformIO $ do
  p <- readIORef counter
  writeIORef counter (p+1)
  return p
\end{code}
The function is certainly not pure since each evaluation of
|new_label ()| (should) yield a new result.
\begin{Exercise}
Why do we need `NOINLINE'?
\end{Exercise}

Smart constructors again hide the labeling. Now they have pure types, 
yielding DSL expressions |ExpL| themselves
rather than expression computations |ExpM|:
\begin{code}
varU :: String -> ExpL
varU v = VariableL (new_label ()) v

constU :: Int -> ExpL
constU x = ConstantL (new_label ()) x

addU :: ExpL -> ExpL -> ExpL
addU x y = AddL (new_label ()) x y
\end{code}

The sample expressions look quite like those in \S\ref{s:cse-problem},
with no traces of labeling, with no leaking of implementation details
into the domain-specific abstraction:
\begin{code}
expU_a = addU (constU 10) (varU "i1")
expU_b = addU expU_a (varU "i2")
\end{code}

The multiplication example also looks just like the pure version in
\S\ref{s:cse-problem}, defining multiplication with familiar
mathematical equations.
\begin{code}
mulU :: Int -> ExpL -> ExpL
mulU 0 _ = constU 0
mulU 1 x = x
mulU n x | n `mod` 2 == 0 = mulU (n `div` 2) (addU x x)
mulU n x = addU x (mulU (n-1) x)
\end{code}
The result of the multiplication by four:
\begin{code}
expU_mul4 = mulU 4 (varU "i1")
\end{code}
shows that identically-named expressions have identical labels and are
hence clearly shared
\begin{code}
AddL 0 (AddL 1 (VariableL 2 "i1") (VariableL 2 "i1")) 
       (AddL 1 (VariableL 2 "i1") (VariableL 2 "i1"))
\end{code}
One look at the labels is enough to see the sharing.

Breaking the referential transparency and lying to the compiler about
effects of our functions may come to haunt us however:
\begin{Exercise}
Why cannot we $\eta$-reduce the smart constructors as follows?
\begin{code}
constU = ConstantL (new_label ())
varU   = VariableL (new_label ())
addU   = AddL (new_label ())
\end{code}
Hint: try it.
\end{Exercise}
Observable sharing will no longer be used in this paper.

\section{Pure Hash-Consing}
\label{s:hash-cons}

Hash-consing is a well-established technique to identify and share
structurally equal data \cite{filliatre-type-safe}. Although the name
comes from Lisp, the technique has been first described prior to Lisp,
in 1957 \cite{Ershov-hash-consing}. The
technique relies on the global mutable hash table mapping structured
values to integer hashes, which are quick to compare. Value
constructors check the table to see if the equal value has been
constructed already, returning the found hash. We describe
hash-consing for a DSL embedded in pure, safe Haskell2010. The
imperative details of hash-consing are hidden better in a
final-tagless style of the DSL embedding, described next.  The
complete code for this section is in the file \url{ExpF.hs} in the
accompanying code.

\subsection{Tagless-final embedding}
\label{s:tagless-final}

In the tagless-final approach \cite{carette-finally-jfp}, embedded DSL
expressions are built with `constructor functions' such as |constant|,
|variable|, |add| rather than the data constructors |Constant|,
|Variable|, |Add| that we have seen in \S\ref{s:cse-problem}. The
constructor functions yield a representation for the DSL expression
being built. The representation could be a string (for
pretty-printing), an integer (for evaluator), etc. Since the same DSL
expression may be concretely represented in several ways, the
constructor functions are polymorphic, parameterized by the
representation |repr|. In other words, the constructor functions are
the members of the type class
\begin{code}
class Exp repr where
  constant :: Int -> repr Int
  variable :: String -> repr Int
  add      :: repr Int -> repr Int -> repr Int
\end{code}
The apparent difference from the datatype |Exp| of
\S\ref{s:cse-problem} is the lower case of the `constructors'. We have
parameterized the representation by expression's type, as common
\cite{carette-finally-jfp}. We did not have to, since the
type so far has been the same, |Int|. The parameterization by the type
will come handy once we add boolean expressions.

The sample expressions from \S\ref{s:cse-problem} look almost the same:
\begin{code}
exp_a = add (constant 10) (variable "i1")
exp_b = add exp_a (variable "i2")
\end{code}
differing only in the lower case of the `constructors'.

The datatype |Exp| from \S\ref{s:cse-problem} is one concrete
(so-called `initial') representation of the DSL expressions~-- one
instantiation of |repr|:
\begin{code}
newtype ExpI t = ExpI Exp

instance Exp ExpI where
  constant = ExpI . Constant
  variable = ExpI . Variable
  add (ExpI x) (ExpI y) = ExpI (Add x y)
\end{code}
\begin{Exercise}
Why do we need the wrapper |ExpI|?
\end{Exercise}

Interpreting |repr| as |ExpI| lets us pretty-print final-tagless
expressions, thanks to the derived |Show| instance for the data type
|Exp|:
\begin{code}
test_shb = case exp_b of ExpI e -> e
-- Add (Add (Constant 10) (Variable "i1")) (Variable "i2")
\end{code}

The multiplication example is largely unchanged, modulo the lower-case
of the constructors and the type signature:
\begin{code}
mul :: Exp repr => Int -> repr Int -> repr Int
mul 0 _ = constant 0
/+\vspace{-15pt}$\cdots$+/

exp_mul4 = mul 4 (variable "i1")
\end{code}

The conversion to |ExpI| took the form of an instance of the class
|Exp| providing the interpretation for the expression primitives, as
the values of the domain |ExpI|. We may write other interpretations,
for example, the evaluator, interpreting an expression as an element
of the domain |R|
\begin{code}
type REnv = [(String,Int)]
newtype R t = R{unR :: REnv -> t} -- A reader Monad
\end{code}
that is, an integer in the environment giving the values for
the free variables occurring in the expression.
\begin{code}
instance Exp R where
  constant x = R (\_ -> x)
  variable x = R (\env -> maybe (error $ "no var: " ++ x) id $ lookup x env)
  add e1 e2  = R (\env -> unR e1 env + unR e2 env)
\end{code}
The evaluator lets us test the multiplication-by-4 example:
\begin{code}
test_val4 = unR exp_mul4 [("i1",5)] -- 20
\end{code}

\begin{Exercise}
Add subtraction and negation to the language. Can we get by without
changing the type class |Exp|, that is, without breaking the existing
code?
\end{Exercise}

\subsection{Detecting implicit sharing}

Recall that our goal is to detect structurally equal subexpressions
and share them, converting an expression tree into a DAG. The
goal is closer if we construct an expression as a DAG to start with. We
represent the DAG as a collection of |Node|s identified by |NodeId|s,
which link the nodes:
\begin{code}
type NodeId = Int
data Node = NConst Int
	  | NVar   String
	  | NAdd   NodeId NodeId
	    deriving (Eq,Ord,Show)
\end{code}
The |Node| data type resembles |Exp| from \S\ref{s:cse-problem};
however, |Node| is not a recursive data type and can be compared in
constant time. The mapping between |Node|s and |NodeId|s is realized
through a |BiMap| interface:
\begin{code}
data BiMap a -- abstract
lookup_key :: Ord a => a -> BiMap a -> Maybe Int
lookup_val :: Int -> BiMap a -> a
insert     :: Ord a => a -> BiMap a -> (Int, BiMap a)
empty      :: BiMap a
\end{code}
|BiMap a| establishes a bijection between the values of the type |a|
and integers, with the operations to retrieve the value given its key,
to find the key for the existing value, and to extend the bijection
with a new association. The type |a| should at least permit equality
comparison; in the present implementation, we require |a| to be a
member of Ord. |BiMap|s can be pretty-printed. Our DAG thus is as
follows:
\begin{code}
newtype DAG = DAG (BiMap Node) deriving Show
\end{code}

Having settled on the DAG implementation we now describe its construction,
which is easier bottom-up. As we construct a node for a subexpression,
we check if the DAG already has the equal node. If so, we return
its |NodeId|; otherwise, we add the node to the DAG. This procedure is
nothing but hash-consing. In fact, it is quite close to Ershov's
original description of hash-consing \cite{Ershov-hash-consing}; our
DAG representation is also similar to his. \aside{See \S2 of
  Ershov's paper; Ershov's CN correspond to our NodeId.} The |BiMap|
interface has exactly the right operations, to check for the presence
of a node and to insert the node, allocating a new |NodeId|. Since
the DAG is being modified as new nodes are built, the construction 
procedure is the State monad computation with the DAG as the
state.

The bottom-up DAG construction maps well to computing a representation
for a tagless-final expression, which is also evaluated bottom-up. The
DAG construction can therefore be written as a tagless-final
interpreter, an instance of the type class |Exp|. The interpreter maps
a tagless-final expression to the concrete representation that is a
|NodeId| in the current DAG:
\begin{code}
newtype N t = N{unN :: State DAG NodeId}

run_expN :: N t -> (NodeId, DAG)
run_expN (N m) = runState m (DAG empty)
\end{code}
The function |run_expN| runs the DAG-construction interpreter and
returns the node, as a reference within a DAG. The construction
algorithm is codified as follows
\begin{code}
instance Exp N where
  constant x = N(hashcons $ NConst x)
  variable x = N(hashcons $ NVar x)
  add e1 e2  = N(do
                 h1 <- unN e1
                 h2 <- unN e2
                 hashcons $ NAdd h1 h2)
\end{code}
with the auxiliary |hashcons| doing the hash-consing, inserting the
|Node| in the DAG if it has not been there already.
\begin{code}
hashcons :: Node -> State DAG NodeId
hashcons e = do
  DAG m <- get
  case lookup_key e m of
    Nothing -> let (k,m') = insert e m
               in put (DAG m') >> return k
    Just k  -> return k
\end{code}

In \S\ref{s:tagless-final} we have defined sample tagless-final
expressions |exp_mul4| and |exp_mul8| for the multiplication by 4 and 8
and interpreted them in several ways, as an integer value and an
expression tree. We now interpret the very same expressions as DAGs:
|run_expN exp_mul4| produces the result
\begin{code}
(2,DAG BiMap[(0,NVar "i1"),(1,NAdd 0 0),(2,NAdd 1 1)])
\end{code}
whereas |run_expN exp_mul8| gives
\begin{code}
(3,DAG BiMap[(0,NVar "i1"),(1,NAdd 0 0),(2,NAdd 1 1),(3,NAdd 2 2)])
\end{code}
A DAG is printed as the list of |(NodeId,Node)| associations. The
sharing of the left and right summands is patent,

The shown results are in fact netlists: a low-level representation of
a circuit listing the gates and their connections, used in circuit
manufacturing. Since our |BiMap| allocated monotonically increasing
|NodeId|s, the resulting netlist comes out topologically sorted.
Therefore, we can straightforwardly generate machine code after the
standard register allocation.

\begin{Exercise}
The method |add| in the |Exp N| instance looks quite like
the function |addM| in Exercise~\ref{ex:addM}. The |addM| function
was flawed. Why does |add| work?
\end{Exercise}

The imperative nature of hash-consing is well-hidden behind the pure
final-tagless interface. We stress this point with the |sklansky|
example, of computing the running sum of several expressions. 
The function |sklansky| was defined in \S\ref{s:cse-problem} with the signature
|sklansky :: (a -> a -> a) -> [a] -> [a]|. In particular,
|sklansky Add| applied to the list of four variables produced the
following list of (pretty-printed) expressions
\begin{code}
["v1","(v1+v2)","((v1+v2)+v3)","((v1+v2)+(v3+v4))"]
\end{code}

We will use the very same |sklansky| to build a `DAG
forest': the list of nodes sharing children within \emph{the same}
DAG. The result is emphatically not the list of isolated
DAGs. Rather, we return the list of |NodeId|s all referring to the
same DAG structure, sharing the components not only within the same
expression but also across independent expressions. The result might
seem impossible yet is easily achievable: we build the list of
DAG-constructing \emph{computations} and then run them in
|sequence|, with the same DAG state:
\begin{code}
test_sklansky n = runState sk (DAG empty)
  where
  sk = sequence (map unN (sklansky add xs))
  xs = map (variable . show) [1..n]
\end{code}
Until the very end, the monadic nature of the DAG construction was
hidden. We manipulated tagless-final expressions as pure, mapping and
passing them around without any regard for their possible effects.
Sharing will be detected nevertheless. The running sum for the list
of four variables, |test_sklansky 4|, now reads
\begin{code}
([0,2,4,7],
  DAG BiMap[
    (0,NVar "1"),(1,NVar "2"),(2,NAdd 0 1),
    (3,NVar "3"),(4,NAdd 2 3),
    (5,NVar "4"),(6,NAdd 3 5),
    (7,NAdd 2 6)])
\end{code}
The repeated expression |v1+v2|, represented by |NodeId|
|2|, is built only once and referenced at three places.

\begin{Exercise}
Our current implementation of |BiMap| relies on a pair of finite maps. We
could have used a highly optimized hash table from the Haskell standard
library. However, hash table operations are performed in the |IO|
monad. Can we accommodate such mutable hash tables without leaking the
|IO| monad, maintaining the form of the |Exp N| interpreter and the
purity of tagless-final expressions?
\end{Exercise}

We have demonstrated the sharing detection technique that represents
a DSL program as a DAG, eliminating multiple occurrences of common
subexpressions. Alas, to find all these common subexpressions we have
to examine the entire expression tree, which may take long time for large
programs (large circuits). The next section describes this problem 
and its solution by explicit sharing.

\section{Explicit sharing}
\label{s:obj-let}

This section motivates the extension of the DSL with a syntactic form
to explicitly indicate expression sharing, and describes its
implementation. This form lets the programmer state their view of a
computation as `common', to be executed once and its result
shared. The programmer thus helps the DSL compiler as well as the human
readers of the code.

The case for the sharing form as part of the pure, non-imperative DSL is
compelling but subtle. At first blush, the
host language |let| form seems sufficient~-- and it is, for some eDSL
interpreters. The tagless-final DSL embedding helps clarify the
subtlety. Our running example will be the familiar
multiplication-by-4, written explicitly below:\footnote{
The complete code for this section is in the file \url{ExpLet.hs}.}
\begin{code}
exp_mul4 = 
 let x = variable "i1" in
 let y = add x x in
 add y y
\end{code}
The two occurrences of the variable |y| refer to the same
tagless-final expression (namely, |add x x|). Such a representation of
a repeated expression by a variable makes the code compact. 
One may also expect that in the internal representation of |exp_mul4|,
the two arguments of |add| refer to the same run-time object.
\begin{Exercise}
Is there any guarantee, in the Haskell Report or the GHC documentation
that two occurrences of the same variable refer to the common shared
object rather than duplicated objects?
\end{Exercise}
Recall that |add x x| is a Haskell computation producing a particular
representation of the DSL expression. What is shared in |exp_mul4| is
the computation rather than the representation. This sharing of
computations, along with the memoization inherent in GHC, speeds up
DSL interpretations. It appears therefore that the Haskell |let| is
sufficient to express explicit sharing: it makes sharing easy to
see in the code and it speeds up interpretations.  The following two
tagless-final interpreters show that the Haskell |let| may indeed
speed up some interpretations, but not the others.

The first interpreter computes the size of an expression, in
the number of its constructors:
\begin{code}
newtype Size t = Size Int
instance Exp Size where
    constant _ = Size 1
    variable _ = Size 1
    add (Size x) (Size y) = Size (x+y+1)
\end{code}
The computed size of |exp_mul4| is |7|. When the computation |add x x|
is first referred to from |y|, the computation will be performed and
its result memoized. The second reference to the same computation via
|y| will use the already determined result. The call-by-need evaluation
strategy of GHC performs shared computations only once. Therefore,
computing the size of even |mul (2^30) (variable "i1")| is
instantaneous. Although that Haskell expression denotes a large DSL
expression tree, the computation over the tree is compactly
represented and is fast to perform.

The other interpreter prints a DSL expression
\begin{code}
newtype Print t = Print (IO ())
instance Exp Print where
    constant = Print . putStr . show 
    variable = Print . putStr
    add (Print x) (Print y) = Print (x >> putStr " + " >> y)
\end{code}
Printing |exp_mul4| gives |i1 + i1 + i1 + i1|, with the subexpression
|i1+i1| duplicated. The duplication is no surprise since our DSL has no
sharing form and hence no way to indicate the sharing in the
print-out. We stress that the printing of |i1+i1| was done
two times. As before, the computation to print |add x x| was shared, and
yet it has evidently been performed twice.
\begin{Exercise}
Why the |Size t| computations were memoized but |Print t| computations
apparently were not? Is there is something special about |IO|?
What about other monads, such as |State|, implemented
as pure functions?
\end{Exercise}
The |Print| interpreter will therefore take a long time to print
|mul (2^30) (variable "i1")|, even if we redirect
the output to |/dev/null|. Thus for some DSL interpreters
including the DAG-constructing interpreter and almost any other
DSL compiler, the running time will be at least proportional
to the size of the DSL expression rather to the size of
the Haskell code to construct the expression. The compactly written
Haskell code may represent exponentially large DSL expressions.

Compacting DSL expressions themselves requires a sharing form in
the DSL itself. We call the form |let_| and add it to our
DSL by defining the type class |ExpLet|.  (Tagless-final embedded
DSLs are extended by introducing a new type class that describes the
syntax of the new syntactic form.  Such a change does not break the
existing expressions written in the non-extended DSL.)
\begin{code}
class ExpLet repr where
  let_ :: repr a -> (repr a -> repr b) -> repr b
\end{code}
The |let_| of DSL, like the |let| of Haskell, expresses sharing
through a local variable binding; multiple occurrences of the
|let_|-bound variable within the |let_| body all refer to the same
expression, the one to which the variable is bound. The DAG
constructing interpretation of |let_|, described below, will make
clear that a |let_|-bound variable refers to the result of the shared
expression. The form |let_| is a binding form, and is embedded in
Haskell using the higher-order abstract syntax, with Haskell's
$\lambda$-bound variable representing the DSL local variable. As an
example of |let_|, we re-write |exp_mul4| indicating the sharing
explicitly:
\begin{code}
exp_mul4' = 
 let_ (variable "i1") (\x ->
 let_ (add x x)       (\y->
 add y y))
\end{code}

We tell the existing tagless-final interpreters how to deal with
|let_|. For example, the |R| interpreter treats |let_| as the flipped
application:
\begin{code}
instance ExpLet R where
  let_ x f = f x
\end{code}
The evaluation of the sample expressions
\begin{code}
val_mul4  = unR exp_mul4  [("i1",5)] -- 20
val_mul4' = unR exp_mul4' [("i1",5)] -- 20
\end{code}
shows that |exp_mul4| with and without explicit sharing evaluate to
the same results. After all, sharing (of pure expressions) is an
optimization and should not affect the results of DSL programs.
\begin{Exercise}
Extend the |Size| interpreter to account for explicit sharing
(that is, write the instance |ExpLet Size|). The size of shared
expressions should be counted only once.
\end{Exercise}

To `see' the sharing, we need a |show|-like function, or an
interpreter of tagless-final expressions as strings. Just strings will
not suffice: to show sharing as let-expressions we need to generate
local variable names. The domain of the show-interpretation is hence
|S t|:
\begin{code}
type LetVarCount = Int
newtype S t = S{unS :: LetVarCount -> String}
\end{code}
\begin{Exercise}
Write the tagless-final interpreter for |S t|, that is, 
the instances |Exp S| and |ExpLet S|.
\end{Exercise}
The |S| interpreter shows |exp_mul4| as
\begin{code}
i1 + i1 + i1 + i1
\end{code}
and |exp_mul4'| as
\begin{code}
let v0 = i1 in let v1 = v0 + v0 in v1 + v1
\end{code}

We tell the DAG-constructing interpreter |N| how to handle explicit
sharing.  Recall that the expression |let_ e (\x -> body)| states that
multiple occurrences of the variable |x| in the |body| should refer to
the same shared expression |e|. In the |N| interpreter, the meaning of
a DSL expression is the DAG-constructing computation producing a
|NodeId|. Sharing a computation across several places means performing
the computation once and replicating its result. This principle is
codified as follows
\begin{code}
instance ExpLet N where
  let_ e f = N(do
               x <- unN e
	       unN $ f (N (return x)))
\end{code}
The result of interpreting |exp_mul4'| as a DAG is identical to that
of |exp_mul4|: the two expressions are indeed identical after the
common subexpression elimination. In |exp_mul4'|, sharing was
explicitly declared; in |exp_mul4| is had to be determined. The
explicit sharing declaration makes the difference in the resources
spent to get the results rather than in the results themselves.

Larger examples will show the difference in the resources. To
obtain the examples, we re-write the |mul| generator to use the
explicit sharing.  The difference from |mul| is on the last-but-one
line.
\begin{code}
mul' :: (ExpLet repr, Exp repr) => Int -> repr Int -> repr Int
mul' 0 _ = constant 0
mul' 1 x = x
mul' n x | n `mod` 2 == 0 = let_ x (\x' -> mul' (n `div` 2) (add x' x'))
mul' n x = add x (mul' (n-1) x)
\end{code}
\begin{Exercise}
There is some sharing left to discover, isn't there?
Modify |mul'| to explicitly declare all sharing.
\end{Exercise}
Without explicit sharing, running the DAG construction 
|run_expN (mul n (variable "i"))| in GHCi takes 0.09 secs for
|n| equal to $2^{12}$, and 0.20 secs for |n| equal to $2^{13}$.
With explicit sharing, running of the
|run_expN (mul' n (variable "i"))| takes the same 0.01 secs
for |n| equal to $2^{12}$, or $2^{20}$ or even $2^{30}$. The
construction is so fast that its timing is lost in noise, even for
the expression with $2^{31}-1$ constructors (most of which are
fortunately shared).

The programmer does not have to explicitly declare all sharing. Some
amount of sharing could be left implicit, for the DAG constructor to
discover. The declared sharing may significantly speed up the
detection of the implicit sharing. For example, the |mul'| code did
not explicitly declared all sharing, as seen from the printout of
|(mul' 15 (variable "i"))|:
\begin{code}
i + let v0 = i in v0 + v0 + 
      let v1 = v0 + v0 in v1 + v1 + let v2 = v1 + v1 in v2 + v2
\end{code}
The DAG construction will find the undeclared sharing, producing the DAG
\begin{code}
(6,DAG BiMap[
   (0,NVar "i"),
   (1,NAdd 0 0), (2,NAdd 1 1), (3,NAdd 2 2),
   (4,NAdd 2 3), (5,NAdd 1 4), (6,NAdd 0 5)])
\end{code}
For a large example, |run_expN (mul (2^30-1) (variable "i"))| finishes
within the same 0.01 secs (although producing the twice as large DAG). The
explicit sharing helps find the remaining implicit sharing.

\begin{Exercise}
How to re-write the |sklansky| example with the explicit sharing?
\end{Exercise}

\section{Conclusions and further reading}
\label{s:concl}

We have demonstrated the sharing detection technique based on the
tagless-final embedding that interprets a DSL program as a DAG,
eliminating multiple occurrences of common subexpressions.  We have
argued for the extension of the DSL with the syntactic form |let_| 
to declare sharing explicitly. The sharing declarations not
only help the human readers of the code but also reliably, sometimes
exponentially, speed up DSL interpreters. Implicit (not stated but
detected) and explicit sharing play well together: the programmer does
not have to identify all expressions to share; the declared sharing
helps, often significantly, to detect the implicit one.

The technique, illustrated on arithmetic expressions, is immediately
applicable to hardware description eDSLs. The technique has also been
used in a SAT solver~\cite{funsat} and in the audio
synthesizer mescaline~\cite{mescaline}.

The standard, thorough reference for compiling embedded DSLs is
Elliott et al. \cite{elliott-compiling}. Sections 4 and 8.1 of the
paper discuss the detection and representation of sharing, with the
particular attention to the placement of the target-code
|let|-expressions to state the sharing in the target code. In the
presence of loops and conditionals, the semantics-preserving
|let|-insertion is quite non-trivial, as the paper discusses in
detail. To transform an expression tree to a DAG, the paper relied on
``non-declarative pointer manipulation'', or so-called ``observable
sharing'', which we illustrated in \S\ref{s:ptr-cmp}.  Broadly,
observable sharing denotes any use of |unsafePerformIO| for the
detection of sharing \cite{Naylor-obs-sharing}. Gill
\cite{Gill-observable} has demonstrated that sharing at certain types is
observable, in the |IO| monad. However, we have to resort to
GHC-specific |StableName|s  and accept their unreliability. \S12 of
\cite{Gill-observable} describes the advantages and many precautions of
observable sharing. 

Detecting sharing is crucial in hardware description languages, since
the modeled circuits are general graphs rather than trees.  The
concise review of the long history of representing sharing in hardware
description embedded DSLs is given in \S2.4.1 of the thesis
\cite{acosta-hardware}.  The thesis describes in more detail the
approaches we have touched upon in \S\ref{s:ptr-cmp}.

Generating code with |let|-expressions to show sharing also has long
history.  The subtle aspects and the need for writing the generator in
the continuation-passing (or monadic) styles or using control effects
have been observed long time ago in partial evaluation community
\cite{bondorf-improving}. See \cite[\S3.1]{carette-multi-stage-scp}
for the detailed explanation of the problem specifically in the
context of code generation.

\begin{Exercise}
Add recursion or iteration to our DSL. Do we need to extend the DSL
with a new syntactic form (e.g., |loop|), or recursive definitions of
Haskell will suffice?
\end{Exercise}

\begin{Exercise}
Add boolean expressions: |true| and |false| literals, conjunctions and
disjunctions, integer comparison. Statically detect errors like taking
the disjunction of integers.
\end{Exercise}

\begin{Exercise}
Add the conditional operator to the DSL. Should control-flow be taken
into account when searching for common subexpressions and sharing them?
\end{Exercise}

\subsubsection*{Acknowledgement}
I am indebted to Chung-chieh Shan for encouragement and many helpful
discussions.


\nocite{*}
\bibliographystyle{eptcs}
\bibliography{sharing}

\begin{thebibliography}{10}
\providecommand{\bibitemdeclare}[2]{}
\providecommand{\urlprefix}{Available at }
\providecommand{\url}[1]{\texttt{#1}}
\providecommand{\href}[2]{\texttt{#2}}
\providecommand{\urlalt}[2]{\href{#1}{#2}}
\providecommand{\doi}[1]{doi:\urlalt{http://dx.doi.org/#1}{#1}}
\providecommand{\bibinfo}[2]{#2}

\bibitemdeclare{mastersthesis}{acosta-hardware}
\bibitem{acosta-hardware}
\bibinfo{author}{Alfonso Acosta-G{\'o}mez} (\bibinfo{year}{2007}):
  \emph{\bibinfo{title}{Hardware Synthesis in {F}or{S}y{D}e}}.
\newblock Master's thesis, \bibinfo{school}{Dept.\ of Microelectronics and
  Information Technology, Royal Institute of Technology},
  \bibinfo{address}{Stockholm, Sweden}.

\bibitemdeclare{inproceedings}{bondorf-improving}
\bibitem{bondorf-improving}
\bibinfo{author}{Anders Bondorf} (\bibinfo{year}{1992}):
  \emph{\bibinfo{title}{Improving Binding Times Without Explicit
  {CPS}-Conversion}}.
\newblock In \bibinfo{editor}{Clinger}  \cite{lfp1992}, pp.
  \bibinfo{pages}{1--10}.

\bibitemdeclare{misc}{funsat}
\bibitem{funsat}
\bibinfo{author}{Denis Bueno} (\bibinfo{year}{2009}):
  \emph{\bibinfo{title}{funsat-0.6.0: A modern {DPLL}-style {SAT} solver}}.
\newblock
  \bibinfo{howpublished}{\url{http://hackage.haskell.org/package/funsat-0.6.0}
  \url{http://hackage.haskell.org/packages/archive/funsat/0.6.0/doc/html/Funsa%
t-Circuit.html}}.

\bibitemdeclare{article}{carette-multi-stage-scp}
\bibitem{carette-multi-stage-scp}
\bibinfo{author}{Jacques Carette} \& \bibinfo{author}{Oleg Kiselyov}
  (\bibinfo{year}{2011}): \emph{\bibinfo{title}{Multi-stage Programming with
  Functors and Monads: Eliminating Abstraction Overhead from Generic Code}}.
\newblock {\sl \bibinfo{journal}{Science of Computer Programming}}
  \bibinfo{volume}{76}(\bibinfo{number}{5}), pp. \bibinfo{pages}{349--375}.

\bibitemdeclare{article}{carette-finally-jfp}
\bibitem{carette-finally-jfp}
\bibinfo{author}{Jacques Carette}, \bibinfo{author}{Oleg Kiselyov} \&
  \bibinfo{author}{Chung-chieh Shan} (\bibinfo{year}{2009}):
  \emph{\bibinfo{title}{Finally Tagless, Partially Evaluated: Tagless Staged
  Interpreters for Simpler Typed Languages}}.
\newblock {\sl \bibinfo{journal}{Journal of Functional Programming}}
  \bibinfo{volume}{19}(\bibinfo{number}{5}), pp. \bibinfo{pages}{509--543},
  \doi{10.1017/S0956796809007205}.

\bibitemdeclare{inproceedings}{claessen-observable}
\bibitem{claessen-observable}
\bibinfo{author}{Koen Claessen} \& \bibinfo{author}{David Sands}
  (\bibinfo{year}{1999}): \emph{\bibinfo{title}{Observable Sharing for
  Functional Circuit Description}}.
\newblock In \bibinfo{editor}{Thiagarajan} \& \bibinfo{editor}{Yap}
  \cite{asian1999}, \doi{10.1007/3-540-46674-6\_7}.

\bibitemdeclare{proceedings}{lfp1992}
\bibitem{lfp1992}
\bibinfo{editor}{William~D. Clinger}, editor (\bibinfo{year}{1992}):
  \emph{\bibinfo{title}{Proceedings of the 1992 {ACM} Conference on {L}isp and
  Functional Programming}}. {\sl \bibinfo{series}{Lisp Pointers}}
  \bibinfo{volume}{V(1)}, \bibinfo{publisher}{{ACM} {P}ress},
  \bibinfo{address}{{N}ew {Y}ork}.

\bibitemdeclare{article}{elliott-compiling}
\bibitem{elliott-compiling}
\bibinfo{author}{Conal Elliott}, \bibinfo{author}{Sigbjorn Finne} \&
  \bibinfo{author}{Oege de~Moor} (\bibinfo{year}{2003}):
  \emph{\bibinfo{title}{Compiling Embedded Languages}}.
\newblock {\sl \bibinfo{journal}{Journal of Functional Programming}}
  \bibinfo{volume}{13}(\bibinfo{number}{3}), pp. \bibinfo{pages}{455--481},
  \doi{10.1017/S0956796802004574}.

\bibitemdeclare{article}{Ershov-hash-consing}
\bibitem{Ershov-hash-consing}
\bibinfo{author}{A.~P. Ershov} (\bibinfo{year}{1958}): \emph{\bibinfo{title}{On
  programming of arithmetic operations}}.
\newblock {\sl \bibinfo{journal}{Communications of the {ACM}}}
  \bibinfo{volume}{1}(\bibinfo{number}{8}), pp. \bibinfo{pages}{3--6},
  \doi{10.1145/368892.368907}.

\bibitemdeclare{inproceedings}{filliatre-type-safe}
\bibitem{filliatre-type-safe}
\bibinfo{author}{Jean-Christophe Filli{\^a}tre} \& \bibinfo{author}{Sylvain
  Conchon} (\bibinfo{year}{2006}): \emph{\bibinfo{title}{Type-Safe Modular
  Hash-Consing}}.
\newblock In {\relax ML}  \cite{ml2006}, pp. \bibinfo{pages}{12--19},
  \doi{10.1145/1159876.1159880}.

\bibitemdeclare{inproceedings}{Gill-observable}
\bibitem{Gill-observable}
\bibinfo{author}{Andy Gill} (\bibinfo{year}{2009}):
  \emph{\bibinfo{title}{Type-safe observable sharing in {H}askell}}.
\newblock In \bibinfo{editor}{Weirich}  \cite{haskell2009}, pp.
  \bibinfo{pages}{117--128}, \doi{10.1145/1596638.1596653}.

\bibitemdeclare{misc}{Hawkins-CSE}
\bibitem{Hawkins-CSE}
\bibinfo{author}{Tom Hawkins} (\bibinfo{year}{2008}): \emph{\bibinfo{title}{I
  love purity, but it's killing me.}}
\newblock
  \bibinfo{howpublished}{\url{http://www.haskell.org/pipermail/haskell-cafe/20%
08-February/039339.html}}.

\bibitemdeclare{misc}{mescaline}
\bibitem{mescaline}
 (\bibinfo{year}{2010}): \emph{\bibinfo{title}{Mescaline: a data-driven audio
  sequencer and synthesizer}}.
\newblock \bibinfo{howpublished}{\url{http://mescaline.puesnada.es/}
  \url{http://mescaline.puesnada.es/doc/html/mescaline/src/Mescaline-Synth-Pat%
tern-AST.html}}.

\bibitemdeclare{proceedings}{ml2006}
\bibitem{ml2006}
 (\bibinfo{year}{2006}): \emph{\bibinfo{title}{2006 {ACM} {SIG\-PLAN} Workshop
  on {ML}}}. \bibinfo{publisher}{{ACM} {P}ress}, \bibinfo{address}{{N}ew
  {Y}ork}.

\bibitemdeclare{misc}{Naylor-sharing}
\bibitem{Naylor-sharing}
\bibinfo{author}{Matthew Naylor} (\bibinfo{year}{2008}):
  \emph{\bibinfo{title}{Designing DSL with explicit sharing}}.
\newblock
  \bibinfo{howpublished}{\url{http://www.haskell.org/pipermail/haskell-cafe/20%
08-February/039671.html}}.

\bibitemdeclare{misc}{Naylor-obs-sharing}
\bibitem{Naylor-obs-sharing}
\bibinfo{author}{Matthew Naylor} (\bibinfo{year}{2008}):
  \emph{\bibinfo{title}{I love purity, but it's killing me.}}
\newblock
  \bibinfo{howpublished}{\url{http://www.haskell.org/pipermail/haskell-cafe/20%
08-February/039347.html}
  \url{http://www.haskell.org/pipermail/haskell-cafe/2008-February/039449.html%
}}.

\bibitemdeclare{inproceedings}{ODonnell-dagstuhl}
\bibitem{ODonnell-dagstuhl}
\bibinfo{author}{John~T. O'Donnell} (\bibinfo{year}{2003}):
  \emph{\bibinfo{title}{Embedding a {H}ardware {D}escription {L}anguage in
  {T}emplate {H}askell}}.
\newblock In \bibinfo{editor}{Christian Lengauer}, \bibinfo{editor}{Don~S.
  Batory}, \bibinfo{editor}{Charles Consel} \& \bibinfo{editor}{Martin
  Odersky}, editors: {\sl \bibinfo{booktitle}{Domain-Specific Program
  Generation}}, {\sl \bibinfo{series}{{L}ecture {N}otes in {C}omputer
  {S}cience}} \bibinfo{volume}{3016}, \bibinfo{publisher}{Springer}, pp.
  \bibinfo{pages}{143--164}, \doi{10.1007/978-3-540-25935-0\_9}.

\bibitemdeclare{proceedings}{asian1999}
\bibitem{asian1999}
\bibinfo{editor}{P.~S. Thiagarajan} \& \bibinfo{editor}{Roland H.~C. Yap},
  editors (\bibinfo{year}{1999}): \emph{\bibinfo{title}{{A}sian {C}omputing
  {S}cience {C}onference}}. {\sl \bibinfo{series}{{L}ecture {N}otes in
  {C}omputer {S}cience}} \bibinfo{volume}{1742}.

\bibitemdeclare{misc}{Thielemann}
\bibitem{Thielemann}
\bibinfo{author}{Henning Thielemann} (\bibinfo{year}{2008}):
  \emph{\bibinfo{title}{I love purity, but it's killing me.}}
\newblock
  \bibinfo{howpublished}{\url{http://www.haskell.org/pipermail/haskell-cafe/20%
08-February/039343.html}}.

\bibitemdeclare{proceedings}{haskell2009}
\bibitem{haskell2009}
\bibinfo{editor}{Stephanie Weirich}, editor (\bibinfo{year}{2009}):
  \emph{\bibinfo{title}{Proceedings of the 2nd {ACM} {SIG{\-}PLAN} Symposium on
  {H}askell}}. \bibinfo{publisher}{{ACM} {P}ress}, \bibinfo{address}{{N}ew
  {Y}ork}.

\end{thebibliography}
\end{document}